\documentclass[12pt]{article}
\usepackage{epsfig}
\usepackage{amsfonts}
\usepackage{latexsym}
\usepackage{amsmath}
\usepackage{verbatim}
\usepackage[dvipdfm,hypertex]{hyperref}
\def\half{{1\over 2}}
\numberwithin{equation}{section}

 \def\p{\partial}

\DeclareFontFamily{OT1}{rsfs}{}
\DeclareFontShape{OT1}{rsfs}{m}{n}{
<-7> rsfs5 <7-10> rsfs7 <10-> rsfs10}{}
\DeclareMathAlphabet{\mycal}{OT1}{rsfs}{m}{n}

\newcommand{\bea}{\begin{eqnarray}}
\newcommand{\eea}{\end{eqnarray}}
\newcommand{\be}{\begin{equation}}
\newcommand{\ee}{\end{equation}}

\newcommand{\bigO}{\mathcal{O}}
\newcommand{\bigL}{\mathcal{L}}

\renewcommand{\t}{\theta}
  \makeatletter
  \let\over=\@@over \let\overwithdelims=\@@overwithdelims
  \let\atop=\@@atop \let\atopwithdelims=\@@atopwithdelims
  \let\above=\@@above \let\abovewithdelims=\@@abovewithdelims
  \makeatother
\begin{document}

\begin{titlepage}
\today
\vskip3cm
\begin{center}

 {\Large \bf  Holographic Derivation of Kerr-Newman Scattering Amplitudes for General Charge and Spin
 }

\vskip 1.5 cm

Thomas Hartman, Wei Song, and Andrew Strominger

 \vskip 1.5 cm

{\it Center for the Fundamental Laws of Nature\\
   Harvard University, Cambridge, MA 02138, USA}\\

\vspace{0.5cm}

\end{center}

\vskip 0.6cm

\begin{abstract}

\noindent
Near-superradiant scattering of charged scalars and fermions by a near-extreme Kerr-Newman black hole and photons and gravitons by a near-extreme Kerr black hole are computed as certain Fourier transforms of correlators in a two-dimensional conformal field theory. The results agree with the classic spacetime calculations from the
1970s, thereby providing good evidence for a conjectured Kerr-Newman/CFT correspondence.
\end{abstract}

\vspace{3.0cm}

\end{titlepage}
\pagestyle{plain}
\setcounter{page}{1}
\newcounter{bean}
\baselineskip18pt


\setcounter{tocdepth}{1}
\tableofcontents

\section{Introduction}

It was recently conjectured that near-extreme Kerr black holes are holographically dual to certain two-dimensional (2D) conformal field theories (CFTs) \cite{Guica:2008mu}.
This is known as the Kerr/CFT correspondence.
If correct, this means that all the properties of near-extreme Kerr - classical or quantum - can be derived from a computation in the dual CFT.  The conjecture was motivated by the fact that, given several apparently benign assumptions, a careful analysis of the properties of diffeomorphisms acting near the horizon actually implies that the near-extreme Kerr  microstates are those of a 2D CFT. The analysis further produces the central charge of the CFT as $c=12J/\hbar $ where $J$ is the angular momentum.  The spectrum of the CFT is deduced from the spectrum of elementary particle in nature.

Both quantum and classical  evidence in favor of the conjecture have emerged. On the quantum front, assuming the validity of the  Cardy formula,  the CFT microstate degeneracy reproduces  the macroscopic Bekenstein-Hawking entropy\cite{Guica:2008mu, Matsuo:2009sj, Castro:2009jf}.
On the classical front, scattering of perturbations near the superradiant bound by a near-extreme Kerr black hole can be holographically computed from correlation functions  in the dual CFT. These were found to
reproduce \cite{Bredberg:2009pv} in full detail the rather complicated expressions derived in the early 70s \cite{Teukolsky:1973ha,Starobinsky:1973,StarobinskyAndChurilov:1973,Press:1973zz}.

While near-extreme Kerr black holes are of direct and significant astrophysical interest,
in order to better understand the structure of the duality it is also of interest to consider more general types of black holes embellished by extra charges, fields  and dimensions. Comparisons of gravity and CFT computations for  these more general objects have in all cases corroborated (generalizations of) the proposed correspondence \cite{gen}, including for Kerr-Newman\cite{Hartman:2008pb}.  This universality is expected because, at heart, the correspondence rests simply on properties of diffeomorphisms. In this paper, we consider the scattering of charged scalars and fermions
from a near-extreme Kerr-Newman black hole, as well as fields of  spin 1 and 2 by a neutral Kerr black hole.\footnote{The wave equation for arbitrary spin
and charge on Kerr-Newman has not been solved.} Perfect agreement between the CFT and gravity computations is found.

A natural quantity to compute is the absorption probability $P_{abs}$. In the regime of interest, the wave equation is solved in the  near-horizon region and the  ``far" asymptotically flat region and then matched along their common boundary. $P_{abs}$ gets a contribution from each region. In the dual CFT picture, the near region is removed from the spacetime and replaced by a CFT
glued along the boundary. It is therefore the near region contribution alone which we expect to be reproduced by the CFT.  The classical formula for this contribution can be extracted from the early papers \cite{Teukolsky:1973ha,Starobinsky:1973,StarobinskyAndChurilov:1973,Press:1973zz,Teukolsky:1974yv}.
A  massive, charge $e$, spin $s=0,\half$ field with energy $\omega$ and angular momentum $m$ scattered against a Kerr-Newman black hole with mass $M$ and charge $Q$ has near-region absorption probability
\be
P_{abs}^{near} \sim {(T_H)^{2\beta}\left(e^{n \pi} - (-1)^{2s}e^{-n\pi}\right)\over \Gamma(2\beta)^2} |\Gamma(\half + \beta -s + ik)|^2 |\Gamma(\half + \beta + i(n-k))|^2 \ ,
\ee
where
\bea
k &=& 2 M \omega - e Q \\
n &=& {\omega - m \Omega_H - e \Phi\over 2\pi T_H} \ .\notag
\eea
Here $2\pi T_H, \Omega_H, \Phi$ are the surface gravity, angular velocity and electric potential of the horizon and we consider the near-superradiant-bound near-extreme scaling limit $T_H\to 0$ with $n$ fixed. $\beta$, given below,  is related to a separation constant that depends on $s$ and must be determined numerically.  For a massless spin $s= 1,2$ field scattered against a Kerr black hole, exactly the same formula applies, but with $e=Q=\Phi=0$. In this paper we will show that these formulae are all Fourier transforms of the CFT correlation functions, in agreement with the Kerr-Newman/CFT correspondence.

The present paper should be viewed as an extension of \cite{Bredberg:2009pv} which treats only the case of neutral scalar scattering by neutral Kerr, but gives more detailed explanations and discussions. In sections 2 and 3 we review
classical Kerr-Newman geometry and the relation of spacetime scattering amplitudes to CFT correlators.  Section 3, 4 and 5 then match the appropriate spacetime and CFT amplitudes for charged scalars on Kerr-Newman, charged fermions on Kerr-Newman and massless spin one and two on Kerr respectively.
Appendix  A presents the near-horizon limit of the Teukolsky master equation and appendix B treats the generalization to magnetic charges.

As this work was nearing completion we received the preprint \cite{Cvetic:2009jn} which has substantial overlap with section 4, and also contains other generalizations.

\section{Macroscopic geometry}
\subsection{Kerr-Newman geometry}
The metric of a Kerr-Newman black hole with mass $M$, angular momentum $J = aM$, and electric charge $Q$ is
\begin{equation}\label{knmetric}
 ds^2= -  {{\Delta} \over \rho^2}( d\hat{t} - a \sin^2\theta d\hat{\phi})^2 + {\rho^2\over \Delta}d\hat{r}^2 + \rho^2 d\theta^2 + {1\over \rho^2}\sin^2\theta\left(a d\hat{t} - (r^2 + a^2)d\hat{\phi}\right)^2 \ ,
\end{equation}
where
\bea
\Delta &=& (\hat{r}^2+a^2)-2M\hat{r}+Q^2 \ ,\\
\rho^2 &=& \hat{r}^2+a^2\cos^2\theta\  .\notag
\eea
The gauge field and field strength are
\begin{align}\label{FS}
    A &= - \frac{Q \hat{r}}{\rho^2}\left( d\hat{t} - \sin^2\theta d\hat{\phi} \right) , \\
    F &= - \frac{Q(\hat{r}^2 - a^2\cos^2\t)}{\rho^4}
    \left( d\hat{t} - a\sin^2\t d\hat{\phi} \right) \wedge d\hat{r} \notag \\
    & \quad - \frac{2Q\hat{r}a\cos\t}{\rho^4}\sin\t d\t \wedge
    \left( ad\hat{t} - (\hat{r}^2 + a^2) d\hat{\phi}  \right).\notag
\end{align}
There are horizons at
\be
r_\pm = M \pm \sqrt{M^2 - a^2 - Q^2} \ ,
\ee
and the entropy, Hawking temperature, angular velocity of the horizon, and electric potential are
\bea
S &=& {\mbox{Area}\over 4} = \pi (r_+^2 + a^2) \\
T_H  &=& {r_+ - r_-\over 4\pi(r_+^2+ a^2)} \notag\\
\Omega_H &=& \frac{ a}{r_+^2 + a^2}  \notag\\
\Phi &=& {Q r_+ \over r_+^2 + a^2} \ .\notag
\eea
We also define the dimensionless Hawking temperature
\be
\tau_H \equiv {r_+ - r_-\over r_+}  \ .
\ee

\subsection{NHEK-Newman geometry}
The extreme black hole has $r_+ = r_- = M$.  Following \cite{Bardeen:1999px,Hartman:2008pb}, we define new coordinates
\begin{equation}\label{NearHor}
\begin{split}
&\hat{r}=r_+ + \lambda r_0 r\ ,\\
&\hat{t}=t r_0/\lambda\ ,\\
&\hat{\phi}=\phi+\Omega_H\frac{t r_0}{\lambda}\ ,
\end{split}
\end{equation}
with $r_0^2 = r_+^2 + a^2$.  Taking $\lambda \to 0$, the near horizon geometry is
\begin{equation}\label{NHKNA}
 ds^2=\Gamma(\theta)\left[
-r^2dt^2+\frac{dr^2}{r^2}
+ d\theta^2 \right] + \gamma(\theta)(d\phi+b rdt)^2\ ,
\end{equation}
where
\begin{eqnarray}\label{definefuncs}
\Gamma(\theta) &=& r_+^2 + a^2 \cos^2\theta\ \\
\gamma(\theta) &=& \frac{(r_+^2+a^2)^2\sin^2\theta}{r_+^2 +a^2 \cos^2\theta} \notag \\
b &=& {2ar_+\over r_+^2 + a^2} \ . \notag
\end{eqnarray}
The near-horizon isometry group is enhanced to $U(1)_L \times SL(2,R)_R$ generated by
\begin{align}
&K_1 = \p_\phi\ ,\notag
&\bar{K}_1 = \p_t\ ,\qquad
\bar{K}_2 = t \p_t - r \p_r, \qquad
\bar{K}_3 = \left({1\over 2r^2} + {t^2\over 2}\right)\p_t - t r \p_r - {b\over r}\p_\phi\ .
\end{align}
Below, in the discussion of spinors, we will use the Newman-Penrose formalism \cite{Newman:1961qr}, which involves a null tetrad $e_a^\mu = (l^\mu, n^\mu, m^\mu, \bar{m}^\mu)$. $m^\mu$ is complex with $\bar{m} = m^*$, and the non-vanishing inner products are
\be
l \cdot n = -m \cdot \bar{m} = -1 \ .
\ee
Slightly generalizing the near-horizon tetrad of \cite{Amsel:2009ev}, in the basis $(t,r,\theta,\phi)$ we use
\bea\label{nhektetrad}
l^\mu &=& \left({1\over r^2}, 1, 0, -{b\over r}\right) \\
n^\mu &=& {1\over 2\Gamma(\theta)}(1, -r^2, 0, -br) \notag\\
m^\mu &=& {1\over \sqrt{2}}\left(0,0, {-i\over \rho_\theta^*}, {\rho_\theta \over \sqrt{\gamma(\theta)\Gamma(\theta)}}\right) \ , \notag
\eea
where
\be
\rho_\theta  =r_+ + i a \cos\theta \ .
\ee

\subsection{Extremal thermodynamics}
The first law of thermodynamics for Kerr-Newman is
\be
T_H dS = dM - \Omega_H dJ - \Phi dQ \ .
\ee
At extremality, $T_H = 0$, so extremal variations satisfy $dM = \Omega_H dJ + \Phi dQ$.  In this case the first law reads
\be\label{extremalfirstlaw}
dS = {1\over T_L}\left(dJ - \mu_L dQ\right) \ ,
\ee
where \cite{Hartman:2008pb}
\bea\label{knpotentials}
T_L &=& {r_+^2 + a^2\over 4 \pi J}\\
\mu_L &=& - {Q^3 \over 2 J} \ .\notag
\eea
According to the Kerr/CFT correspondence, $T_L$ is the left-moving temperature of the dual 2d CFT.

\section{Microscopic scattering }
We will consider the scattering of various fields off the Kerr and Kerr-Newman black holes.  According to the Kerr-Newman/CFT correspondence, a  bulk field $\Psi$ is dual to a CFT operator $\mathcal{O}$, and the scattering cross section for $\Psi$ is related to the CFT two-point function \cite{Maldacena:1997ih,Bredberg:2009pv}
\be\label{cftt}
G(t^+,t^-) = \langle \bigO^\dagger(t^+,t^-)\bigO(0)\rangle \ ,
\ee
where $t^\pm$ are the coordinates of the 2d CFT. From Fermi's golden rule, the absorption cross section is
\be\label{cftform}
P_{abs} \sim \int dt^+dt^- e^{-i\omega_R t^- - i\omega_Lt^+}\left[G(t^+-i\epsilon,t^--i\epsilon) - G(t^++i\epsilon,t^-+i\epsilon)\right]  \ .
\ee
At left and right temperatures $(T_L, T_R)$ in chemical potentials $(\mu_L, \mu_R)$, an operator with conformal dimensions $(h_L, h_R)$ and charges $(q_L, q_R)$ has two-point function
\be\label{gzerotemp}
G \sim (-1)^{h_L+h_R}\left(\pi T_L\over \sinh(\pi T_L t^+)\right)^{2h_L} \left(\pi T_R\over \sinh(\pi T_R t^-)\right)^{2h_R}e^{iq_L \mu_L t^+ +iq_R\mu_Rt^-}  \ ,
\ee
determined by conformal invariance.
Performing the integral in (\ref{cftform}),
\bea\label{cftsigma}
P_{abs}
&\sim& T_L^{2h_L-1}T_R^{2h_R-1} \left(e^{\pi \tilde{\omega}_L + \pi \tilde{\omega}_R} \pm e^{-\pi \tilde{\omega}_L - \pi \tilde{\omega}_R}\right) |\Gamma(h_L + i \tilde{\omega}_L) |^2 |\Gamma(h_R + i \tilde{\omega}_R) |^2  \ ,
\eea
where
\be\label{cftfreq}
\tilde{\omega}_L = {\omega_L - q_L \mu_L\over 2 \pi T_L} \ , \quad \ \tilde{\omega}_R = {\omega_R - q_R \mu_R\over 2\pi T_R} \ .
\ee
The two-point function (\ref{cftt}) has a branch cut, and as a result, one must find a way to fix the choice of relative sign between the two exponentials in (\ref{cftsigma}).  While there may be a way to do this from first principles we will simply fix the sign by matching the computations. The result (\ref{cftsigma}) is easily generalized to include more chemical potentials by further shifts in the frequencies (\ref{cftfreq}).

We will refer back to (\ref{cftsigma}) throughout the paper to compare our bulk computations to the CFT result under various circumstances.  In order to make the comparison, we must specify the temperatures and chemical potentials, and for each field the conformal weights, charges, and momenta $(\omega_L, \omega_R)$.

\section{Charged scalar}
We first consider a scalar field
\be
\Psi = e^{-i\omega \hat{t} + i m \hat{\phi}}R_0(\hat{r})S_0^\ell(\theta) \ ,
\ee
with charge $e$ and mass $\mu$ in the Kerr-Newman geometry (\ref{knmetric}). The case $Q = e = 0$ was considered in \cite{Bredberg:2009pv}. The generalization to include magnetic charges is given in appendix B. The wave equation separates into the angular part
\be\label{scalarangular}
\left[{1\over \sin\theta}\p_\theta(\sin\theta \p_\theta) +K_{\ell} - a^2(\omega^2 - \mu^2)\sin^2\theta - {m^2\over \sin^2\theta} \right] S_0^\ell(\theta) = 0 \ ,
\ee
and the radial part
\be\label{scalarradial}
\p_r(\Delta \p_r R_0) + V_0 R_0 = 0
\ee
with
\bea\label{scalarradialb}
V_0 &=& -K_\ell + 2 a m \omega + {H^2\over \Delta}-\mu^2(\hat{r}^2 + a^2) \\
H &=& \omega(r^2 + a^2) - e Q r - a m \ .\notag
\eea
$K_\ell$ is a separation constant determined numerically by regularity at $\theta = 0,\pi$. Defining
\bea\label{definek}
x &=& {\hat{r}-r_+\over r_+} \\
n &=& {\omega - m \Omega_H - e \Phi\over 2 \pi T_H} \notag\\
k &=& 2 r_+ \omega - e Q \ ,\notag
\eea
this becomes
\be\label{scalarradialnew}
x(x +\tau_H)R'' + (2x +\tau_H)R' + V_0 R = 0
\ee
with
\bea\label{scalarradialpot}
V_0 &=& -K_\ell + 2am\omega + {H^2\over r_+^2 x(x+\tau_H)}-\mu^2\left(r_+^2(x+1)^2+a^2\right)\\
H &=& r_+^2 \omega x^2 + r_+ k x + n \tau_H r_+/2 \ . \notag
\eea
We will work in the regime with
\be\label{regime}
\tau_H \ll 1  \ , \quad M(\omega - m\Omega_H - e\Phi) \ll 1
\ee
with $n$ finite. That is, we consider fields with momentum near the superradiant bound $\omega \sim m \Omega_H + e \Phi$ scattered by near-extreme black holes.

\subsection{Near region}
With $x \ll 1$, the wave equation is
\be
x(x+\tau_H)R'' + (2x+\tau_H)R' + V_{near} R  = 0
\ee
with
\be
V_{near} = -K_\ell + 2am\omega + {(kx+n\tau_H/2)^2\over x(x+\tau_H)}   - \mu^2(r_+^2 + a^2)\ .
\ee
This is the wave equation on the near horizon geometry in thermal coordinates.  It is a special case of the NHEK master wave equation solved in the appendix.  The solution ingoing at the horizon is
\be\label{rinnear}
R_{near}^{in} = x^{-{i\over 2}n}\left({x\over \tau_H}+1\right)^{{i\over 2}n - i k}\,_2F_1\left(\half +\beta -ik, \half - \beta -ik, 1 - in, -{x\over \tau_H}\right)
\ee
where
\be\label{scalarbeta}
\beta^2 = K_\ell + {1\over 4} - 2 am\omega -k^2  + \mu^2(r_+^2 +a^2)\ .
\ee
Since we are working in the regime (\ref{regime}) to leading order in $\tau_H$, here $\beta$ and $k$ can be evaluated at extremality and at the superradiant bound $\omega = m\Omega_H + e \Phi$. We will only consider the case of real $\beta>0$. (Imaginary $\beta$ modes require more care, as in \cite{Bredberg:2009pv}.) For $x \gg \tau_H$,
\be\label{scalarbdry}
R_{near}^{in} = A x^{-\half + \beta} + B x^{-\half - \beta}
\ee
with
\bea
A &=& \tau_H^{\half - \beta-in/2}{\Gamma(2\beta)\Gamma(1-in)\over\Gamma(\half+\beta-ik)\Gamma(\half+\beta-i(n-k))} \\
B &=& \tau_H^{\half + \beta-in/2}{\Gamma(-2\beta)\Gamma(1-in)\over\Gamma(\half-\beta-ik)\Gamma(\half-\beta-i(n-k))} \ .\notag
\eea
Note that for real $\beta$ and $\tau_H \ll 1$,
\be
B \ll A \ .
\ee

\subsection{Scattering amplitude}
The full scattering cross section can be computed easily by solving the wave equation in the far region $x \gg \tau_H$ and matching to $R_{near}^{in}$. However, we will need only the near horizon contribution in order to match to the CFT.  From the full absorption probability
\be
P_{abs} = {\mathcal{F}_{abs}\over \mathcal{F}_{in}} \ ,
\ee
the near horizon contribution is defined by stripping off the magnitude of the source at the boundary $x=x_B$,
\be
P_{abs}^{near} = {P_{abs}\over |\Psi(x=x_B)|^2}
\ee
where $\Psi$ is normalized to have unit incoming flux and $\tau_H \ll x_B \ll 1$.  The normalization can also be accounted for by using the manifestly near-region formula
\be
P_{abs}^{near} = {\mathcal{F}_{abs} \over |\Psi(x=x_B)|^2 }  \ .
\ee
The wavefunction (\ref{rinnear}) is normalized to have unit flux at the horizon, so using $B \ll A$,
\bea\label{scalarkn}
P_{abs}^{near} &\sim& {1\over |A|^2}\notag\\
&\sim& {\tau_H^{2\beta}\sinh(\pi n) \over \Gamma(2\beta)^2}|\Gamma(\half + \beta + i k)|^2|\Gamma(\half + \beta + i(n-k))|^2 \ .
\eea

\subsection{Conformal weight}\label{s:scalarweight}
The boundary value of the field $\Psi$ acts as a source for a CFT operator $\bigO$ with left and right conformal dimensions
\be
L_0 = h_L \ , \quad \bar{L}_0 = h_R \ .
\ee
$h_R$ is determined by the $SL(2,R)_R$ isometry of the near horizon geometry.  In \cite{Bredberg:2009pv}, $h_R$ for a scalar was derived by organizing solutions to the near horizon wave equation into representations of the isometry group. Here we will use a different  argument, which is quicker because we have already solved the wave equation.

On the near horizon geometry (\ref{NHKNA}), the zero mode of $SL(2,R)_R$ is
\be
\bar{L}_0 = t \p_t - r \p_r \  .
\ee
This generates the scale transformation
\be\label{scaletrans}
t \to \zeta t \ , \quad r \to \zeta^{-1} r \ .
\ee
From (\ref{scalarbdry}), the leading behavior of a scalar near the boundary is
\be\label{scalfall}
\Phi \sim \Phi_0(t,\phi,\theta)r^{-\half + \beta} \ ,
\ee
so under the scale transformation,
\be
\Phi_0 \to \Phi_0 \zeta^{\half - \beta} \ .
\ee
Therefore conformal invariance implies that $\Phi_0$ can act as the source for a boundary operator with scaling dimension
\be\label{scaldim}
h_R = \half + \beta \ .
\ee

\subsection{Comparison to CFT}\label{scalarcftcomp}
We can now compare the gravity result (\ref{scalarkn}) to the CFT result (\ref{cftsigma}).  To relate the two, we take
\bea\label{scalarqn}
h_L &=& h_R = \half + \beta \ ,\\
\omega_L &=& m \ , \quad \ T_L = {M^2 + a^2\over 4 \pi J} \ ,  \notag\\
\mu_L &=& - {Q^3\over 2 J} \ , \quad  \ q_L = e \ ,\notag \\
\mu_R &=& \Omega_H \ , \quad \ q_R = m \ .\notag
\eea
$h_R$ was derived above, and $h_L = h_R$ is the natural choice for a scalar.  $\omega_L$ was derived in \cite{Guica:2008mu,Bredberg:2009pv},  $T_L$ and $\mu_L$ were derived in (\ref{knpotentials}), and since $\mu_L$ is the electric potential, $q_L = e$.  Finally, the right-moving temperature and quantum number are defined by equating the near-horizon and asymptotic Boltzmann factors \cite{Guica:2008mu}
\be\label{nsplit}
n = {\omega - m \Omega_H - e \Phi\over 2 \pi T_H}  = {\omega_R - q_R \mu_R\over 2\pi T_R} + {\omega_L - q_L \mu_L\over 2\pi T_L} \ .
\ee

To leading order, the quantity $k$ defined in (\ref{definek}) that appears in the gravity result can be written
\bea
k &=& 2 r_+ \omega - e Q\\
&=& {m - e \mu_L\over 2\pi T_L} \notag\\
&=& \tilde{\omega}_L \ ,\notag
\eea
and from (\ref{nsplit}),
\be
n - k = \tilde{\omega}_R \ .
\ee

Putting this all together, and choosing the undetermined relative sign in (\ref{cftsigma}) to be negative, the gravity and CFT results agree.

\section{Charged fermion}
We now consider a Dirac fermion $\psi$ with charge $e$ and mass $\mu$ scattered by a Kerr-Newman black hole.  The wave equation on Kerr was separated by Chandrasekhar \cite{Chandrasekhar:1976ap,Chandrasekhar:1985kt} and extended to Kerr-Newman by Page \cite{Page:1976jj}.  Writing $\psi = (P_A, \bar{Q}^{A'})^T$, the Dirac equation is
\bea
\sqrt{2}(\nabla_{BB'} - i e A_{BB'})P^B + i \mu Q^*_{B'} &=& 0 \\
\sqrt{2}(\nabla_{BB'} + i e A_{BB'})Q^B + i \mu P^*_{B'} &=& 0 \ .\notag
\eea
Write the wavefunctions
\bea\label{fermionwavefunc}
\psi &=& \left(-P^1, P^0, \bar{Q}^{0'}, \bar{Q}^{1'}\right) \\
&=& e^{-i\omega\hat{t} + i m \hat{\phi}}\left(- R_\half S_\half,{R_{-\half}S_{-\half}\over \sqrt{2}(\hat{r}-ia\cos\theta)}, -{R_{-\half}S_\half\over \sqrt{2}(\hat{r} + i a\cos\theta)}, R_\half S_{-\half}\right)  \notag
\eea
where $R_s = R_s(\hat{r})$ and $S_s = S_s^\ell(\theta)$. Defining
\be
\bigL_s \equiv \p_\theta + 2 s (m \csc\theta -  a \omega \sin\theta) + \half \cot\theta \ ,
\ee
the angular equations are
\be\label{chargedangular}
\left[\bigL_{-s}{1\over \Lambda_{\ell} - 2 s a \mu \cos\theta}\bigL_s + \Lambda_\ell + 2 s a \mu\cos\theta\right]S_s^\ell(\theta) = 0 \ ,
\ee
where $\Lambda_\ell$ is a separation constant.
The radial equation is
\be
\Delta^{-s}\p_{\hat{r}}\left(\Delta^{s+1}\p_{\hat{r}} R_s\right) + {2is\mu\Delta\over \Lambda_\ell - 2i s \mu \hat{r}}\p_{\hat{r}} R_s + V_s R_s = 0
\ee
with
\bea
V_s &=& {H^2 - 2 i s(\hat{r}-M)H\over \Delta} + 2s(s+\half){\Lambda_\ell - i M\mu\over \Lambda_\ell - 2s i \mu \hat{r}}\\
& &  \ \ \ + 4 i s \omega \hat{r} - 2 i s e Q - {\mu H\over \Lambda_\ell - 2 i s \mu \hat{r}}  - \mu^2 \hat{r}^2 - \Lambda_\ell^2 \notag \ .
\eea
This can be rewritten
\be
x(x+\tau_H)R_s'' + (1+s) (2x + \tau_H)R_s'+ {2isr_+ x \mu(x+\tau_H)\over \Lambda_\ell - 2isr_+(1+x)\mu}R_s' + V_s R_s = 0 \ .
\ee
The relative normalization of the radial components is determined by
\be\label{relnorm}
R_{\half} = {1\over \Lambda_\ell+i\mu\hat{r}}\left(\p_{\hat{r}} - {i H\over\Delta}\right)R_{-\half} \ .
\ee

\subsection{Near region}
When $x \ll 1$, in the regime (\ref{regime}), the radial equation is
\be
x(x+\tau_H)R_s'' + (1+s) (2x + \tau_H)R_s' + V_s^{near} R_s = 0
\ee
with
\be
V_s^{near} = {(kx + n\tau_H/2)^2 - is(2x+\tau_H)(kx + n\tau_H/2)\over x(x+\tau_H)} + s(2s+1) + 2 i s k - \mu^2 r_+^2 - \Lambda_\ell^2 \ .
\ee
From the appendix, the ingoing solution is
\bea\label{fermionradialanswer}
R_{s} &=& N_{s} x^{-i{n\over 2} -s}\left({x\over\tau_H}+1\right)^{-s + i({n\over 2}-k)} \\
& & \ \ _2F_1\left(\half + \beta -s -ik, \half - \beta - s -ik, 1-s-in, -{x\over \tau_H}\right)\notag
\eea
where
\be\label{fermionbeta}
\beta^2 = \Lambda_\ell^2 - k^2 + r_+^2\mu^2 \ .
\ee
The relative normalization $N_{\half}/N_{-\half}$ is fixed by (\ref{relnorm}),
\be\label{relnormresult}
{N_{\half}\over N_{-\half}} ={1-2in\over 2r_+(\Lambda_\ell + i \mu r_+)} \ .
\ee

\subsection{Scattering amplitude}
With the fermion flux defined by \cite{Martellini:1977qf,Iyer:1978du}
\be
\mathcal{F} = \Delta|R_\half|^2 - |R_{-\half}|^2 \ ,
\ee
the absorption probability is
\be
P_{abs} = {\mathcal{F}_{abs}\over \mathcal{F}_{in}}  \ .
\ee
(A positive overall constant in $\mathcal{F}$ has been absorbed into the normalization of $S_{\pm \half}$.)
As for scalars, we can extract the near horizon contribution to the absorption probability without solving the far region wave equation:
\be
P_{abs}^{near} \sim {\mathcal{F}_{abs} \over |\Psi(x_B)|^2} \ .
\ee
$\Psi$ is the source at the boundary of the near horizon region,
\be
 \tau_H \ll x_B \ll 1 \ .
\ee
We choose the leading coefficient of either $R_{\half}$ or $R_{-\half}$ as the source.  Due to (\ref{relnormresult}), they have the same magnitude near the boundary, so any linear combination gives the same answer. Computing the absorbed flux at the horizon where $R_{-\half} \to 0$, we find
\bea\label{fermionkn}
P_{abs}^{near} &\sim& { |\sqrt{\Delta}R_{\half}(0)|^2\over |R_{\pm\half}(x_B)|^2} \notag \\
 &\sim& {\tau_H^{2\beta} \cosh (\pi n)\over \Gamma(2\beta)^2} \left|\Gamma( \beta + i k)\right|^2 |\Gamma({1\over 2} + \beta + i (n-k))|^2 \ .
\eea

\subsection{Conformal weight}
We will now determine the right-moving conformal dimension $h_R$ of the spinor operator $\mathcal{O}$ dual to a bulk fermion. As in Section \ref{s:scalarweight}, we need the scaling of the wavefunction $\psi$ near the boundary under
\be
\bar{L}_0 = t\p_t - r\p_r \ .
\ee
Expanding (\ref{fermionradialanswer}) for $x \gg \tau_H$, we see that a fermion in the near horizon geometry (\ref{NHKNA}) behaves near the boundary as
\be\label{boundaryferm}
\psi \sim e^{-i\omega_{near} t + i m \phi}\left(r^{-1 +\beta}S_{\half}(\theta), r^\beta S_{-\half}(\theta), r^\beta S_{\half}(\theta), r^{-1 +\beta} S_{-\half}(\theta)\right) \ ,
\ee
where $\omega_{near}$ comes from the coordinate transformation to the near horizon.  Since we only need the scaling behavior, all relative coefficients have been dropped.

The Lie derivative of a fermion along a Killing vector $\xi$ is \be
\mathcal{L}_\xi \psi =  \xi^\mu \nabla_\mu \psi - {1\over
4}\gamma^{\mu\nu}\nabla_\mu \xi_\nu \psi\ , \ee where $\nabla_\mu
\psi = \left(\p_\mu + {1\over 4}\omega_{\mu ab}\gamma^{ab}\right)
\psi$ with $\omega_{\mu \ b}^{\ a}$ the spin connection.  The gamma
matrices are \be\label{gammas} \gamma^\mu = \sqrt{2}\left(
               \begin{array}{cc}
                 0 & \sigma^\mu_{AB'} \\
                 \bar{\sigma}^{\mu A'B} & 0 \\
               \end{array}
             \right) \ , \quad
\sigma^\mu_{AB'} =  \left(
                      \begin{array}{cc}
                        l^\mu & m^\mu \\
                        \bar{m}^\mu & n^\mu \\
                      \end{array}
                    \right) \ ,
\ee
with $\bar{\sigma}^\mu = -\epsilon \sigma^{\mu T} \epsilon$, $\epsilon_{01} = 1$. The Newman-Penrose tetrad $(l,n,m,\bar{m})$ was given in (\ref{nhektetrad}). Setting $\xi = \bar{L}_0$ and using (\ref{boundaryferm}), we find near the boundary
\be
\mathcal{L}_\xi \psi = (\half - \beta - i \omega_{near} t)\psi \ .
\ee
Therefore the boundary value of $\psi$ is a source for a CFT operator of dimension
\be
h_R = \half + \beta \ .
\ee
\subsection{Comparison to CFT}
We now compare the gravity result (\ref{fermionkn}) to the general CFT scattering amplitude (\ref{cftsigma}).  For the left- and right-moving temperatures, potentials, and quantum numbers, we choose the same identifications as for scalars in (\ref{scalarqn},\ref{nsplit}). The only difference is that now
\be
h_L = \beta \ , \quad \ h_R = \half + \beta \ .
\ee
$h_R$ was derived above, and $|h_R - h_L| = \half$ is natural for fermions. Choosing the undetermined sign in (\ref{cftsigma}) to be a plus, the near horizon contribution to fermion scattering is exactly reproduced by the dual CFT.

\section{Photons and gravitons}
The electromagnetic and gravitational perturbations of Kerr-Newman do not decouple \cite{Chandrasekhar:1985kt,Berti:2009kk}. Therefore in this section we specialize to the uncharged Kerr black hole, for which the problem was solved by Starobinsky and Churilov \cite{Starobinsky:1973,StarobinskyAndChurilov:1973} and Press and Teukolsky \cite{Teukolsky:1973ha,Press:1973zz,Teukolsky:1974yv}. We will review their macroscopic derivation of the scattering amplitude, then compare to the microscopic CFT result.

The radiative fields
\be
\psi_s = e^{-i\omega \hat{t} + i m \hat{\phi}}S_s^{\ell}(\theta)R_s(\hat{r}) \ ,
\ee
which are related to the field strength and Weyl tensor for spin-1 ($s=\pm1$) and spin-2 ($s=\pm2$) perturbations respectively, satisfy the Teukolsky equations
\be\label{teukspin}
{1\over \sin \theta}\p_\theta(\sin\theta \p_\theta S_s^\ell) + \left(K_{\ell}^s - {m^2+s^2+2ms \cos\theta\over \sin^2\theta} - a^2 \omega^2 \sin^2\theta - 2 a \omega s \cos\theta \right) S_s^\ell = 0
\ee
\be\label{teukradial}
\Delta^{-s}\p_{\hat{r}}\left(\Delta^{s+1}\p_{\hat{r}}R_s\right) + \left({H^2-2is(\hat{r}-M)H\over\Delta} + 4is\omega\hat{r} + 2 a m \omega +s(s+1)- K_\ell^s\right)R_s = 0 \ .
\ee
The detailed relation between $\psi_s$ and the field perturbations $\phi$, $A_\mu$, $h_{\mu\nu}$ will be given below in Section \ref{s:fieldpert}. We normalize the angular modes so that
\be
\int d\theta \sin\theta (S_s^{\ell})^2 = 1 \ .
\ee
The relative normalization of $+|s|$ and $-|s|$ radial modes is determined by
\be\label{spinnorma}
\mathcal{D}^{2|s|}R_{-|s|} = {B_{|s|}\over 2^{|s|}} R_{|s|}
\ee
with
\bea\label{spinnormb}
\mathcal{D} &=& \p_{\hat{r}} - iH/\Delta \\
|B_1|^2 &=& (K_\ell^{(1)} - 2 a m \omega)^2 + 4 a m \omega - 4 a^2 \omega^2\notag \\
|B_2|^2 &=& (Q_\ell^2 + 4 a m \omega - 4a^2 \omega^2)\left[(Q_\ell-2)^2 + 36 a m \omega - 36 a^2 \omega^2\right] \notag\\
& & \ \ \ + (2Q_\ell-1)(96a^2\omega^2 - 48 a \omega m) + 144 \omega^2(M^2-a^2) \notag\\
Q_\ell &=& K_\ell^{(2)} - 2 a m \omega \ .\notag \eea In terms of $x
= {\hat{r}-r_+\over r_+}$, the radial equation is
\be\label{radialsx} x(x+\tau_H)R_s'' + (s+1)(2x + \tau_H)R_s' + V_s
R_s = 0 \ee with \bea
V_s &=& { (r_+\omega x^2 + 2 r_+ \omega x + n\tau_H/2)^2 - is(2x + \tau_H)(r_+\omega x^2 + 2 r_+ \omega x + n\tau_H/2)\over x(x+\tau_H)} \\
& & \ \ \ + 4 ir_+ \omega s (1+x) + 2am\omega +s(s+1)- K_\ell^s \ .\notag
\eea

\subsection{Near region}
For $x \ll 1$, the radial equation is (\ref{radialsx}) with
\be
V_s^{near} = { (m x + n\tau_H/2)^2 - is(2x + \tau_H)(m x + n\tau_H/2)\over x(x+\tau_H)} + 2 i m s  + m^2 + s(s+1) - K_\ell^s \ ,
\ee
where we have replaced $2r_+\omega$ and $2 a \omega$ by the leading order value $m$. This is the NHEK master equation considered in the appendix (\ref{nhekmaster}), with ingoing solution
\be\label{spinnear}
R_s^{near} = x^{-i{n\over 2} -s }\left({x\over\tau_H}+1\right)^{i({n\over 2} -m)-s}\ _2F_1\left(\half+\beta-s-im, \half-\beta-s-im,1-s-in,-{x\over \tau_H}\right) \ ,
\ee
where
\be\label{spinbeta}
\beta^2 = {1\over 4} + K_\ell^s - 2m^2 \ .
\ee

\subsection{Far region}
For $x \gg \tau_H$, the radial equation is
\be
x^2 R_s'' + (s+1)2x R_s' + V_s^{far}R_s = 0
\ee
with
\be
V_s^{far} = -K_\ell^s + m^2 + {m^2\over 4}(x+2)^2 + imsx + s(s+1)\ .
\ee
The solution is
\be\label{spinfar}
R_s^{far} = A x^{-\half + \beta - s}e^{-imx/2}\ _1F_1\left(\half + \beta -s+im, 1 + 2\beta, imx\right) + B(\beta \to -\beta )\ .
\ee

\subsection{Matching}
In the matching region $\tau_H \ll x \ll 1$,
\be\label{spinmatching}
R_s^{far} \to Ax^{-\half + \beta - s} + Bx^{-\half -\beta - s} \ .
\ee
Comparing to the large-$x$ expansion of $R_s^{near}$, we find
\bea
A &=& {\Gamma(2\beta)\Gamma(1-s-in)\over\Gamma(\half + \beta -i(n-m))\Gamma(\half + \beta - s - im)}\tau_H^{\half-i{n\over 2} - \beta}\\
B &=& {\Gamma(-2\beta)\Gamma(1-s-in)\over\Gamma(\half - \beta -i(n-m))\Gamma(\half - \beta - s - im)}\tau_H^{\half-i{n\over 2} + \beta} \ .\notag
\eea

\subsection{Scattering}
The absorption probability is the rate of absorbed energy per unit incoming energy,
\be\label{spinpabs}
P_{abs} = {dE_{abs}/dt\over dE_{in}/dt} \ .
\ee
Writing the asymptotic behavior of the field near infinity
\be
R_{+|s|} = Y_{in}x^{-1-im} + \cdots
\ee
and near the horizon
\be
R_{+|s|} =  Y_{abs}x^{-i{n\over 2} - s} + \cdots \ ,
\ee
the absorption probability is
\be
P_{abs} = F_{s} \left| Y_{abs} \over Y_{in}\right|^2\ .
\ee
$F_s$ is a spin-dependent `flux factor' that comes from the energy-momentum tensor used to define (\ref{spinpabs}). The derivation appears in \cite{StarobinskyAndChurilov:1973,Teukolsky:1974yv} and will not be repeated here; the result is
\begin{align}
F_0 &= {n \tau_H \over m} &\mbox{(scalar, $s=0$)}&\\
F_1 &= {m \tau_H \over n}  &\mbox{(photon, $s=1$)}& \notag\\
F_2 &=  {m^3 \tau_H\over n(n^2+1)}&\mbox{(graviton, $s=2$)}& \notag \ .
\end{align}
Reading off $Y_{abs}$ from (\ref{spinnear}) and $Y_{in}$ from the asymptotics of (\ref{spinfar}), the final answer is
\be
P_{abs} = F_s e^{\pi m}(m\tau_H)^{2\beta - 1}m^{2-2s}{ |\Gamma{\half +\beta +i(m-n)}|^2 |\Gamma(\half +\beta -s + im)|^2|\Gamma(\half + \beta + s + im)|^2\over \Gamma(2\beta)^2\Gamma(1+2\beta)^2|\Gamma(1-s+in)|^2 }\ ,
\ee
with the positive $s$ taken for each spin.

Schematically, the near horizon contribution is
\be
P_{abs}^{near} \sim {dE_{abs}/dt \over |\Psi(x=x_B)|^2 }
\ee
where $\Psi$ is the `source term' for a CFT operator at the boundary $x=x_B$.  For scalars and fermions, we took the source to be proportional to the leading part of the wave function.  For $s=1,2$, the radial functions $R_s$ are related to the photon field strength and gravitational Weyl tensor, and the proper definition of $\Psi$ depends on the details of the coupling between bulk and boundary fields.  Here we simply assume that the source is proportional to $R_{\pm s}(x_B)$. Then the near horizon contribution to the absorption probability is\footnote{The fact that this does not depend on whether we pick $R_{+|s|}$ or $R_{-|s|}$ for the source comes from the relative normalization of the two perturbations, determined by (\ref{spinnorma},\ref{spinnormb}).}
\be\label{spinkerr}
P_{abs}^{near} \sim {\tau_H^{2\beta}\sinh(\pi n)\over\Gamma(2\beta)^2}|\Gamma(\half + \beta - s+im)|^2|\Gamma(\half + \beta + i(n-m))|^2 \ .
\ee

\subsection{Field perturbations}\label{s:fieldpert}
We have computed the radial functions $R_s(\hat{r})$, but in order to compare to the CFT we will need the scaling behavior of the actual fields $\phi$, $A_\mu$ and $h_{\mu\nu}$ near the boundary.  For a scalar, the relationship is trivial, $\phi = e^{-i\omega\hat{t} + im \hat{\phi}}R_0(\hat{r})S_0^\ell(\theta)$, but for $|s|=1,2$ the conversion from $R_s$ to the field perturbations is more involved.

Following Teukolsky \cite{Teukolsky:1973ha}, the Newman-Penrose tetrad in Boyer-Lindquist coordinates, in the basis $(\hat{t}, \hat{r}, \theta, \hat{\phi})$, is
\bea
l^\mu &=& \left({\hat{r}^2 + a^2\over \Delta},1,0,{a\over \Delta}\right) \ , \quad n^\mu = {1\over 2(\hat{r}^2 + a^2 \cos^2\theta)}\left(\hat{r}^2 + a^2, -\Delta, 0, a\right) \\
m^\mu &=& {1\over \sqrt{2}(\hat{r} + i a \cos\theta)}\left(ia\sin\theta,0,1, {i\over \sin\theta}\right) \ .\notag
\eea
The Teukolsky wave functions
\be
\psi_s = e^{-i\omega \hat{t} + i m \hat{\phi}}S_s^{\ell}(\theta)R_s(\hat{r})
\ee
are related to the electromagnetic field strength $F_{\mu\nu}$ and the Weyl tensor $C_{\mu\nu\rho\sigma}$ by \cite{Teukolsky:1973ha}
\bea
\psi_1 &=& F_{\mu\nu} l^\mu m^\nu \\
\psi_{-1} &=& (\hat{r} -ia\cos\theta)^2 F_{\mu\nu}m^{\star \mu}n^\nu \notag\\
\psi_2 &=& = C_{\mu\nu\rho\sigma}l^\mu m^\nu l^\rho m^\sigma \notag\\
\psi_{-2} &=& (\hat{r} -ia\cos\theta)^4 C_{\mu\nu\rho\sigma}n^\mu
m^{\star \nu}n^\rho m^{\star \sigma} \ .\notag \eea On Kerr, these
relations were inverted by Chrzanowski \cite{Chrzanowski:1975wv}.
In terms of the Newman-Penrose spin coefficients \be \alpha\ ,
\beta\ , \tau\ , \rho\ , \epsilon\ , \pi \ , \ee and the
differential operators \be D = l^\mu \p_\mu \ , \quad \delta^* =
m^{* \mu}\p_\mu \ , \ee the inversion formulae with our normalizations are \bea
h_{\mu\nu} &=& \big(-l_\mu l_\nu(\delta^* + \alpha + 3 \beta^* - \tau^*)(\delta^* + 4 \beta^* + 3 \tau^*) - m^*_\mu m^*_\nu(D - \rho^* )(D + 3 \rho^* ) \\
& & + l_{(\mu}m^*_{\nu)}\left[(D + \rho - \rho^*)(\delta^* + 4 \beta^* + 3 \tau^*) + (\delta^* + 3 \beta^* - \alpha - \pi - \tau^*)(D + 3 \rho^* )\right]\big)\notag\\
& & \times {4\over B_2} R_{-2}(\hat{r})S_2^\ell(\theta)e^{-i\omega\hat{t} + i m \hat{\phi}}\notag \\
A_\mu &=& - \left(- l_\mu(\delta^* + 2 \beta^* + \tau^*) + m^*_\mu(D
+ \rho^*)\right]{2\over B_1}R_{-1}(\hat{r} )
S_{1}^{\ell}(\theta)e^{-i\omega \hat{t} + i m \hat{\phi}} \ . \notag
\eea From (\ref{spinmatching}), the radial wave functions behave
near the boundary as \be R_s \sim A x^{-\half + \beta - s} + \cdots
. \ee Plugging in the Kerr spin coefficients from
\cite{Teukolsky:1973ha}, we find the leading behavior of the fields
for $\tau_H \ll x \ll 1$ in the basis
$(\hat{t},x,\hat{\phi},\theta)$, \bea
A_\mu &\sim& \mathcal{O}\left(x^{\half + \beta}, x^{-{3\over 2} + \beta}, x^{-\half + \beta}, x^{-\half + \beta}\right) \ , \label{gaugepert}\\
h_{\mu\nu} &\sim& \mathcal{O}
\left(
  \begin{array}{cccc}
    x^{{3\over 2} + \beta} & x^{-\half + \beta} & x^{\half + \beta} & x^{\half + \beta} \\
     & x^{-{5\over 2} + \beta} & x^{-{3\over 2} + \beta} & x^{-{3\over 2} + \beta} \\
     &  & x^{-\half + \beta} & x^{-\half + \beta} \\
     &  &  & x^{-\half + \beta}\\
  \end{array}
\right) \ .\label{metricpert}
\eea
The metric perturbation (\ref{metricpert}) was first derived in \cite{Amsel:2009ev} from a near-horizon standpoint.

\subsection{Conformal weight}
Next we need the right-moving conformal weight $h_R$ of the operator $\mathcal{O}$ dual to a spin-1 or spin-2 field. The derivation is identical to that for scalars in Section \ref{s:scalarweight}, except that we need to account for tensor indices on the source.

Consider a tensor field near the boundary,
\be\label{genfall}
\Psi = \Psi_0(t,\phi,\theta)r^{\alpha}  + \cdots\,
\ee
with tensor indices suppressed. Under the scale transformation (\ref{scaletrans}),
\be
\Psi_0 \to \Psi_0 \zeta^{-\alpha  +d_t - d_r}
\ee
where $d_t$ is the number of $t$ indices and $d_r$ is the number of $r$ indices on the component of $\Psi$ under consideration.  Therefore $\Psi_0$ can act as the source for a boundary operator with scaling dimension
\be\label{gendim}
h_R = {1+\alpha - d_t + d_r} \ .
\ee
Using the field perturbations (\ref{gaugepert}, \ref{metricpert}) in conjunction with (\ref{genfall}, \ref{gendim}), we find that all components give the same scaling behavior
\be
h_R = \half + \beta \ ,
\ee
for $|s|=0,1,2$. (Note, however, that $\beta$ depends implicitly on $s$ through the separation constants.)

\subsection{Comparison to CFT}
The gravity result (\ref{spinkerr}) agrees with the CFT result (\ref{cftsigma}) if we choose
\be
h_R = \half + \beta \ , \quad \ h_L = \half + \beta - |s| \ .
\ee
The temperatures, potentials, and quantum numbers are as for scalars (\ref{scalarqn}, \ref{nsplit}), except in this section we set $e=Q=\Phi=0$, so there is no left-moving potential.  $h_R$ was derived above, and as expected, $|h_R - h_L| = |s|$.

\section*{Acknowledgements}
This work was supported in part by
DOE grant DE-FG02-91ER40654.

\appendix

\section{Near-horizon master equation }
We commonly encounter radial wave equations of the form
\be\label{nhekmaster}
x(x + \tau_H)R'' +(1+s)(2x+\tau_H)R' + VR = 0 \ ,
\ee
with
\be
V = {(ax + b\tau_H)^2-is(2x+\tau_H)(ax+b \tau_H)\over x(x+\tau_H)}-c \ .
\ee
This is the finite temperature NHEK analog of the Teukolsky master equation (\ref{teukradial}) for spin-$s$ perturbations on Kerr.  The solutions are
\begin{align}
R_1 &= x^{ib}\left({x\over\tau_H}+1\right)^{-s +i(b-a)}\ _2F_1\left(\half -\beta + i(2b-a), \half + \beta +i(2b-a), 1 + s + 2ib, -{x\over \tau_H}\right) \notag\\
R_2 &= x^{-ib-s}\left({x\over\tau_H}+1\right)^{-s+i(b-a)}\ _2F_1\left(\half + \beta-s-ia, \half - \beta -s -ia, 1-s-2ib, -{x\over \tau_H}\right)
\end{align}
where
\be
\beta^2 = {1\over 4} + c + 2ais -a^2 + s(s+1) \ .
\ee

\section{Magnetic charge}
In this appendix we compute the scattering of a scalar with electric and magnetic charges $(q_e, q_m)$ against an extreme dyonic Kerr-Newman black hole with charges $(Q_e, Q_m)$. We find that in the CFT, this corresponds to turning on an additional chemical potential for left-movers.

By electromagnetic duality, this process is equivalent to scattering a scalar with charges $(e=\sqrt{q_e^2 + q_m^2},0)$ against a black hole with charges
\be\label{dualityrot}
\left(
  \begin{array}{c}
    Q_e' \\
    Q_m' \\
  \end{array}
\right)
=
{1\over e}\left(
  \begin{array}{c}
    {q_e Q_e + q_m Q_m} \\
    {q_e Q_m - q_m Q_e} \\
  \end{array}
\right) \ .
\ee
The dyonic Kerr-Newman metric is obtained by replacing $Q^2 \to Q_e'^2 + Q_m'^2$ in (\ref{knmetric}), and the first law of thermodynamics at extremality (\ref{extremalfirstlaw}) becomes
\bea
dS &=& {1\over T_L}\left(dJ - \mu_e Q_e' - \mu_m Q_m'\right)\\
\mu_{e,m} &=& - Q_{e,m}' {Q_e'^2 + Q_m'^2\over 2 J} \ , \notag
\eea
with $T_L$ unchanged.

Adding magnetic charge to the black hole, the scattering computation on the gravity side is modified as follows.  The angular part of the scalar wave equation (\ref{scalarangular}) becomes \cite{Semiz:1991kh}
\be
\left[{1\over \sin\theta}\p_\theta(\sin\theta \p_\theta) +K_{\ell} - a^2(\omega^2 - \mu^2)\sin^2\theta - {(m-e Q_m' \cos\theta)^2\over \sin^2\theta} - 2 a \omega e Q_m' \cos\theta\right] S_0^\ell(\theta) = 0 \ ,
\ee
which alters the separation constant $K_\ell$, and therefore the conformal weight $\half + \beta$.  The radial equation is identical to (\ref{definek},\ref{scalarradialnew},\ref{scalarradialpot}) with the replacement
\bea
k &=& 2 r_+ \omega - e Q_e' \\
&=& {1\over 2\pi T_L}\left(m - q_e \mu_e - q_m \mu_m\right)\notag \ ,
\eea
where in the second line we have used the duality rotation (\ref{dualityrot}) to give magnetic charge to the scalar.  Summarizing, the final result for the scattering cross section (\ref{scalarkn}) is identical, but with modified $\beta$ and $k$.

In the CFT, we now have two left-moving chemical potentials $(\mu_e,\mu_m)$ for the charges $(q_e, q_m)$. Following the steps in Section \ref{scalarcftcomp} with this one modification, we then find perfect agreement between gravity and the CFT.

\end{document}